\documentclass[a4paper]{jpconf}
\usepackage{graphicx}

\newcommand{\be}{\begin{equation}}
\newcommand{\ee}{\end{equation}}
\newcommand{\ba}{\begin{array}}
\newcommand{\ea}{\end{array}}

\newcommand{\bea}{\begin{eqnarray*}}
\newcommand{\eea}{\end{eqnarray*}}
\newcommand{\ei}{\end{itemize}}

\newtheorem{theorem}{Theorem}

\newtheorem{example}{Example}
\begin{document}
\title{The theory of contractions of 2D 2nd order quantum superintegrable systems and its relation to the Askey scheme for hypergeometric orthogonal polynomials}

\author{Willard Miller, Jr.}

\address{ School of Mathematics, University of Minnesota,
 Minneapolis, Minnesota,
55455, U.S.A.}

\ead{miller@ima.umn.edu}

\begin{abstract}
 We describe a contraction theory for 2nd order superintegrable systems, showing that all such systems  in 2 dimensions  are limiting cases of a single system: 
the generic 3-parameter potential on the 2-sphere,  $S9$ in our listing. Analogously, all of the quadratic symmetry algebras of these systems can be
 obtained by a sequence of contractions starting from  $S9$. By contracting function space realizations of irreducible representations of the $S9$ algebra 
(which give the structure equations for
 Racah/Wilson polynomials) to  the other  superintegrable systems one  obtains the 
full Askey scheme of orthogonal hypergeometric polynomials.This relates the scheme directly to explicitly solvable quantum mechanical systems. 
Amazingly, all of these contractions of superintegrable systems with potential are uniquely induced
by Wigner Lie algebra contractions of $so(3,{\bf C})$ and $e(2,{\bf C})$. The present  paper concentrates on describing  this intimate link 
between Lie algebra and superintegrable system contractions, with the detailed calculations presented elsewhere. Joint work with E. Kalnins, S. Post, E. Subag and R. Heinonen.
\end{abstract}

\section{Introduction}
 A  quantum superintegrable system  is an integrable Hamiltonian system on an $n$-dimensional Riemannian/pseudo-Riemannian manifold 
with potential: $ H=\Delta_n+V$,  that admits $2n-1$ 
algebraically independent  partial differential operators commuting with $H$, apparently the maximum  possible. 
\[ [H,L_j]=0,\quad  L_{2n-1}=H, \ n=1,2,\cdots, 2n-1.\]
Here, $\Delta_n$ is the Laplace-Beltrami operator on the manifold.
Superintegrability captures the properties of 
quantum Hamiltonian systems that allow the Schr\"odinger eigenvalue problem $H\Psi=E\Psi$ to be solved exactly, analytically and algebraically. 
There is a similar definition of classical superintegrable systems with Hamiltonian ${\cal H}=\sum g^{jk}p_jp_k+V$ on phase space with $2n-1$ functionally independent 
constants of the motion  ${\cal L}_j$ with ${\cal L}_{2n-1}={\cal H}$ and polynomial in the momenta,  definitely the maximum number possible.
 A system is  of order $K$ if the maximum order of the symmetry
 operators $L_j$, other than $H$, (or classically the maximum order of constants of the motion as polynomials) is  $K$. For $n=2$, $K=2$ all systems are known. 
 The symmetry operators of 
each system close under commutation (or under the Poisson bracket)  to generate a 
quadratic algebra, and 
the irreducible representations of the  algebra determine the eigenvalues of $H$ and their multiplicity. Classically we get important information about the orbits 
through algebraic methods alone. Detailed motivation for the study of superintegrable systems, a presentation of the theory and many references can be found in 
\cite{SCQS, superreview}.
 All the 2nd order classical and quantum superintegrable systems are limiting cases of
 a single system: the generic 3-parameter potential on the 2-sphere,  $ S9$ in our listing. Analogously all  quadratic symmetry algebras of these 
systems are contractions of  $ S9$. In the quantum case this system  is 
\be\label{S90} {S9}:\qquad H=\Delta_2+\frac{a_1}{s_1^2}+\frac{a_2}{s_2^2}+\frac{a_3}{s_3^2}, \quad s_1^2+s_2^2+s_3^2=1,\ee
\[L_1= (s_2\partial_{s_3}-s_3\partial_{s_2})^2 +\frac{a_3 s_2^2}{s_3^2}+\frac{a_2 s_3^2}{s_2^2}, \quad  L_2,\ L_3,\ {\rm obtained\ by \ cyclic\ permutation\ of \ indices},\]
\[ H=L_1+L_2+L_3+a_1+a_2+a_3.\]

In the following sections we give brief descriptions of 1st and 2nd order 2D superintegrable systems, both free and with degenerate or nondegenerate potential. 
Every nonfree system is associated with a closed quadratic algebra generated by its symmetries. We state, and prove elsewhere, that a free system extends to a superintegrable system with potential if and only if its symmetries 
generate a closed free  quadratic algebra. We point out that the theory of contractions of  Lie symmetry algebras of constant curvature spaces  is intimately 
associated with superintegrable systems of 
1st order; indeed it appears to have been the motivation for the development of this theory by Wigner and In\"on\"u. 
Then we show for systems on 2D constant curvature spaces how these Lie algebra contractions induce 1) 
contractions of the free quadratic algebras and then 2) induce contractions of the nondegenerate and degenerate  quadratic algebras of systems with potential. 
Next we describe how 
the contractions of the superintegrable systems with potential can induce contractions of models of irreducible representations of the quadratic algebras through the process of `saving' a 
representation. The Askey scheme for hypergeometric orthogonal polynomials emerges as a special subclass of these model contractions. We conclude with some observations.

\section{1st and 2nd order 2D superintegrable systems}

\medskip\noindent
{\bf 1st order systems $K=1$}: In the quantum case these are the (zero-potential)  Laplace-Beltrami eigenvalue equations on constant curvature spaces,
such as  the Euclidean Helmholtz equation $(P_1^2+P_2^2)\Phi=-\lambda^2 \Phi$ 
(or the  Klein-Gordon equation  $(P_1^2-P_2^2)\Phi=-\lambda^2 \Phi$), and  the  Laplace -Beltrami 
eigenvalue equation on the 2-sphere $(J_1^2+J_2^2+J_3^2)\Psi=-j(j+1)\Psi.$ 
The first order symmetries close under commutation to form  the Lie algebras  $e(2,{\bf R})$, $e(1,1)$  or $o(3,{\bf R})$. 
 The eigenspaces of these systems support differential operator models of the irreducible representations of the Lie algebras in which  basis 
eigenfunctions are the spherical 
harmonics ($o(3,{\bf R})$),Bessel functions ($e(2,{\bf R})$) and more complicated special functions \cite{IPSW,KMP}.

It was exactly these 1st order systems which motivated the pioneering work of In\"on\"u and Wigner \cite{Wigner}
on  Lie algebra contractions. While, that paper introduced Lie algebra contractions in general, the motivation and virtually all the examples were of symmetry 
algebras of these systems. 
It was shown that  $o(3,{\bf R})$ contracts to $e(2,{\bf R})$. In the physical space this is 
accomplished by letting the radius of the sphere go to infinity, so that the surface flattens out. Under this limit the Laplace-Beltrami eigenvalue equation
 goes to the Helmholtz equation. 

The following defines so-called natural contractions, \cite{Saletan}, a generalization of Wigner-In\"on\"u contractions.
\medskip\noindent
{\bf Lie algebra contractions}:
Let $(A; [ , ]_A)$, $(B; [ , ]_B)$ be two complex Lie algebras. We say
$B$ is a {\it contraction} of $A$ if for every $\epsilon\in (0; 1]$ there exists a linear invertible
map $t_\epsilon : B\to A$ such that for every $X, Y\in B$, 
\[ \lim_{\epsilon\to 0}t_\epsilon^{-1}[t_\epsilon X,t_\epsilon Y]_A
= [X, Y ]_B.\]
Thus,  as $\epsilon\to 0$ the 1-parameter family of basis transformations can become nonsingular but the structure constants go to a finite limit.

\medskip\noindent Features of Wigner's contraction approach, \cite{Wigner, Talman}: 
\begin{itemize} 
\item `Saving' a representation. Passing through a sequence of irreducible representations of the source symmetry algebra to obtain an irreducible representation of the target algebra in the contraction limit.
\item Simple models of irreducible representations. Finding models on function spaces so that the eigenfunctions of the generators are special functions.
\item Limit relations between special functions, as a result of saving a model representation in the contraction limit.
\item Use of the models to find expansion coefficients relating different special function bases.
\end{itemize}

\medskip\noindent {\bf Free 2nd order superintegrable systems in 2D}: We will apply Wigner's ideas to 2nd order systems in 2D $(2n-1=3)$. We start with the free 
(no potential function) case.
The complex spaces with free Hamiltonians admitting at least three 2nd order symmetries (i.e., three 2nd order Killing tensors) were classified by Koenigs \cite{Koenigs}. They are:
\begin{itemize} \item  The two constant curvature spaces: flat space and the complex 2-sphere. They each admit 6 linearly independent 2nd order symmetries and 3 1st order symmetries,
 \item The four  Darboux spaces, (4 2nd order symmetries and 1 1st order symmetry):
\[ ds^2=4x(dx^2+dy^2),\ ds^2=\frac{x^2+1}{x^2}(dx^2+dy^2),\]
\[ds^2=\frac{e^x+1}{e^{2x}}(dx^2+dy^2),\ ds^2= \frac{2\cos 2x+b}{\sin^2 2x}(dx^2+dy^2),\]
\item Eleven 4-parameter Koenigs spaces (3 2nd order symmetries and no 1st order symmetries). An example is 
\[ ds^2=(\frac{c_1}{x^2+y^2}+\frac{c_2}{x^2}+\frac{c_3}{y^2}+c_4)(dx^2+dy^2).\]
\end{itemize}

\medskip\noindent {\bf 2nd order superintegrable systems (with potential) in 2D}:
All such systems are known. There are  59 and each of the spaces classified by Koenigs admits at least one system. 
However, under the St\"ackel 
transform, an invertible  structure preserving mapping \cite{superreview},  the systems  divide into 12 equivalence classes, each with a representative in a constant curvature space.
Now the symmetry algebra is a quadratic algebra, 
not usually a Lie algebra, and 
the irreducible representations of the quantum  algebra determine the eigenvalues of $H$ and their multiplicity

There are 3 types of superintegrable systems:
 \begin{enumerate}\item Nondegenerate: (3-parameter potential)
\[V({\bf x})= a_1V_{(1)}({\bf x})+a_2V_{(2)}({\bf x})+a_3V_{(3)}({\bf x})+a_4\]
\item  Degenerate: (1-parameter potential) \[V({\bf x})= a_1V_{(1)}({\bf x})+a_2\]
\item  Free: \[ V=a_1.\]
\end{enumerate}
Usually the trivial added constant in each potential is ignored, though it is vital for some purposes.

\medskip\noindent {\bf Nondegenerate systems ($2n-1=3$ generators)}:
The quantum symmetry algebra generated by $H,L_1,L_2$ always closes under commutation. 
Let    $ R=[L_1,L_2]$ be the 3rd order commutator of the generators. Then  
\[ [L_j,R]=
A_1^{(j)}L_1^2+A_2^{(j)}L_2^2+A_3^{(j)}H^2+A_4^{(j)}\{L_1,L_2\}+A_5^{(j)}HL_1+A_6^{(j)}HL_2\]
\[ +A_7^{(j)}L_1+A_8^{(j)}L_2+A_9^{(j)}H+A_{10}^{(j)}\]
\[  R^2 = b_1 L_1^3 + b_2 L_2^3 + b_3 H^3 + b_4 \{L_1^2,L_2\} + b_5 \{L_1,L_2^2\} + b_6
L_1 L_2 L_1 + b_7 L_2 L_1 L_2\]
$$  + b_8 H\{L_1,L_2\} +b_9 H L_1^2 + b_{10} H L_2^2 + b_{11} H^2
L_1 + b_{12} H^2 L_2 + b_{13} L_1^2 + b_{14} L_2^2 + b_{15} \{L_1,L_2\} $$
$$ + b_{16} H L_1 + b_{17} H
L_2 + b_{18} H^2 + b_{19} L_1 + b_{20} L_2 + b_{21} H + b_{22},$$
Here $\{L_j,L_k\}=L_jL_k+L_kL_j$ is the symmetrizer of $L_j$ and $L_k$.
 This structure is an example of a {\it  quadratic algebra}. Here the  $A_i^{(\ell)},b_j$ are constants or polynomials in the parameters $a_k$ of the potential. The exact rules are given in  \cite{KMPost13} and \cite{superreview}.

\medskip\noindent {\bf Degenerate systems $(2n-1=3)$}:
  There are 4 generators: one 1st order $X$ and 3 second order $H,L_1,L_2$.
 \[[X,L_j]=C_1^{(j)}L_1+C_2^{(j)}L_2+C_3^{(j)}H+C_4^{(j)}X^2+C_5^{(j)},\quad j=1,2,\]
\[ [L_1,L_2]=E_1\{L_1,X\} +E_2\{L_2,X\} +E_3HX+E_4X^3+E_5X, \]
Since $2n-1=3$ there must be an identity satisfied by the 4 generators. It is of 4th order:
\[ c_1 L_1^2+c_2L_2^2+c_3H^2+c_4\{L_1,L_2\}+c_5HL_1+c_6HL_2+c_7X^4+c_8\{X^2,L_1\}+c_9\{X^2,L_2\}\]
\[+c_{10}HX^2+c_{11}XL_1X+c_{12}XL_2X+c_{13}L_1+c_{14}L_2+c_{15}H+c_{16}X^2+c_{17}=0\]
Again the  $C_i,E_j,c_\ell$ are constants or polynomials in the parameters $a_k$ of the potential.

The structure of classical quadratic algebras is similar, except no symmetrizers are needed. In \cite{KMSH} it is shown that all of the classical and quantum structure equations for nondegenerate systems can, in fact, be derived from  the  equation for $R^2$, and all degenerate structure equations can be determined to within a constant factor from the 4th order identity.

\medskip\noindent {\bf St\"ackel Equivalence Classes}:
There are  59 types of 2D 2nd order superintegrable systems, on a variety of manifolds  but under the St\"ackel 
transform, an invertible  structure preserving mapping,  they divide into 12 equivalence classes with representatives on flat space and the 2-sphere, 6 with 
nondegenerate 3-parameter potentials \[\{ S9,E1, E2, E3', E8,E10\}\] and 6 with degenerate 1-parameter potentials, \cite{superreview},
\[ \{ S3, E3, E4, E5, E6, E14\}.\]
The notation comes from \cite{KKMP} where all 2nd order 2D superintegrable systems on constant curvature spaces are classified.

\section{Representatives of nondegenerate quantum systems}

\begin{enumerate} \item
$S9$:  Defined in (\ref{S90}).

\medskip\noindent 
{ Structure equations: }
\[[L_i,R]=4\{L_i,L_k\}-4\{L_i,L_j\}- (8+16a_j)L_j + (8+16a_k)L_k+ 8(a_j-a_k),\]
\[R^2=\frac83\{L_1,L_2,L_3\} -(16a_1+12)L_1^2 -(16a_2+12)L_2^2  -(16a_3+12)L_3^2+\]
\[\frac{52}{3}(\{L_1,L_2\}+\{L_2,L_3\}+\{L_3,L_1\})+ \frac13(16+176a_1)L_1
+\frac13(16+176a_2)L_2 + \frac13(16+176a_3)L_3 \]
\[+\frac{32}{3}(a_1+a_2+a_3)
+48(a_1a_2+a_2a_3+a_3a_1)+64a_1a_2a_3,\quad R=[L_1,L_2].\]

\item {$E1$} (Winternitz-Smorodinsky system)
  \[ H=\partial_x^2+\partial_y^2-\omega^2(x^2+y^2)+\frac{b_1}{x^2}+\frac{b_2}{y^2}\]
 Generators:
\[L_1=\partial_x^2-\omega^2x^2+\frac{b_1}{x^2},\
L_2=\partial_y^2-\omega^2y^2+\frac{b_2}{y^2},\
L_3=(x\partial_y-y\partial_x)^2+y^2\frac{b_1}{x^2}+x^2\frac{b_2}{y^2}\] 
{ Structure relations:}
\[
[R,L_1]=8L_1^2 -8HL_1-16\omega^2L_3 +8\omega^2,\]
\[
[R,L_3]=8HL_3-8\{L_1,L_3\} +(16b_1+8)H-16(b_1+b_2+1)L_1,\]
 \[ R^2+\frac83\{L_1,L_1,L_3\}-8H\{L_1,L_3\}+(16b_1+16b_2+\frac{176}{3})L_1^2-16\omega^2L_3^2-(32b_1+\frac{176}{3})HL_1\]
\[+(16b_1+12)H^2+\frac{176}{3}\omega^2L_3+16\omega^2(3 b_1+3 b_2+4b_1b_2+\frac23)=0\]

\item {$E2$}
\[  H=\partial_x^2+\partial_y^2-\omega^2(4x^2+y^2)+b x+\frac{c}{y^2}\]
{Generators:}
\[  { L}_1=\partial_x^2-4\omega^2 x^2+bx,\ 
{ L}_2=\partial_y^2-\omega^2 y^2+\frac{c}{y^2},\ 
{ L}_3=\frac12\{(x\partial_y-y\partial_x), \partial_y\}+y^2(\omega^2 x-\frac{b}{4})+\frac{cx}{y^2}\]
{ Structure equations:}
\[[L_1,R]+2bL_2-16w^2L_3=0,\ [L_3,R]+2L_2^2-4L_1L_2+2bL_3+\omega^2(8c+6)=0,\]
\[{ R}^2=4{ L}_1{\ L}_2^2+16\omega^2 { L}_3^2-2b\{{ L}_2,{ L}_3\}+(12+16c)\omega^2{ L}_1-32w^2 L_2-b^2(c+\frac34)\]
 Here, the algebra generators are  $H,L_1,L_3,\ R=[L_1, L_3]$

\item {$E3'$}
\[ { H}=\partial_x^2+\partial_y^2-\omega^2(x^2+y^2)+c_1 x+c_2y=L_1+L_2\]
{ Generators}:
\[ { L}_1=\partial_x^2-\omega^2 x^2+c_1x,\
{ L}_2=\partial_y^2-\omega^2 y^2+c_2y,\
{ L}_3=\partial_{xy}-\omega^2 xy+\frac{c_2x+c_1y}{2}\]
{ Structure relations:}
\[{ [L_1,R]}=4\omega^2 L_3-c_1c_2,\ [L_3,R]=-2\omega^2 L_1+2\omega^2 L_2+\frac12(c_1^2-c_2^2),\]
\[{ R}^2=4\omega^2({ L}_3^2-{ L}_1{ L}_2)-2c_1c_2{ L}_3+c_2^2{ L}_1+c_1^2{ L}_2+4\omega^4\]
 The algebra generators are $H, L_1, L_3,\ R=[L_1,L_3]$.
\item {$E10$}
  \[ { H}=\partial_x^2+\partial_y^2+\alpha{\bar z}+\beta(z-\frac32 {\bar z}^2)+\gamma(z{\bar z}-\frac12 {\bar z}^3)\]
{ Generators:} 
\[{ L}_1=(\partial_x-i\partial_y)^2+\gamma {\bar z}^2+2\beta {\bar z},\]
\[{ L}_2=2i\{x\partial_y-y\partial_x, \partial_x-i\partial_y\}+(\partial_x+i\partial_y)^2-4\beta z{\bar z}-\gamma z{\bar z}^2-2\beta {\bar z}^3-
\frac34\gamma{\bar z}^4+\gamma z^2+\alpha {\bar z}^2+2\alpha z\]
{ Structure equations:}
\[[R,L_1]+32\gamma L_1+32\beta^2=0,\quad [R,L_2]-96L_1^2-64\beta H+128\alpha L_1-32\gamma L_2-32\alpha^2,\]
\[ R^2= 
 64{ L}_1^3-64\gamma { H}^2-128\alpha{ L}_1^2+128\beta{ H}{ L}_1+32\gamma\{ L_1, L_2\}-128\alpha\beta{ H}+64\alpha^2{ L}_1+64\beta^2{ L}_2-256\gamma^2.\]
 Here $R=[L_1,L_2]$, {  $z=x+iy$, ${\bar z}=x-iy$},  
\item {$E8$}
 \[ H=\partial_x^2+\partial_y^2+\frac{c_1z}{{\bar z}^3}+\frac{c_2}{{\bar z}^2}+c_3z{\bar z}\]
{ Generators:}
\[{ L}_1=(\partial_x-i\partial_y)^2-\frac{c_1}{{\bar z}^2}+c_3{\bar z}^2,\ 
L_2=(x\partial_y-y\partial_x)^2+c_1\frac{z^2}{{\bar z}^2}+c_2\frac{z}{{\bar z}}\]
{ Structure relations:}
\[[R,L_1]
=8L_1^2+32c_1c_3,\   [R,L_2]=-8\{L_1,L_2\}+8c_2H-16L_1,\]
\[R^2=-\frac{16}{3}\{L_1^2,L_2\} -\frac{16}{3}L_1L_2L_1-\frac{176}{3}L_1^2+16c_1H^2+16c_2L_1H-64c_1c_3L_2
+16c_3(\frac{4}{3}c_1-c_2^2).\]
 Here, $R=[L_1,L_2]$, { $z=x+iy$, ${\bar z}=x-iy$}, 
\end{enumerate}

\section{Representatives of degenerate systems}
There are close relations  between nondegenerate and degenerate systems.
 \begin{itemize}
  \item  Every 1-parameter potential can be obtained from some 3-parameter
 potential by  parameter restriction. 
\item It is not simply a restriction, however, because the structure of the symmetry algebra changes. 
\item A  formally skew-adjoint 1st order symmetry appears and this induces a new 2nd order symmetry. 
\item Thus the restricted potential has a strictly larger symmetry algebra  than is initially apparent.
 \end{itemize}
We list the 6 representatives of the equivalence classes for degenerate systems:
\begin{enumerate} \item {$S3$ (Higgs Oscillator)} 
\[ H=J_1^2+J_2^2+J_3^2+\frac{a}{s_3^2}\]
The system is the same as $S9$ with $a_1=a_2=0$, $a_3=a$ 
with  the former $L_2$ replaced by 
\[
 L_2=\frac12(J_1J_2+J_2J_1)-\frac{a s_1s_2}{s_3^2}\]
  and  \[ X=J_3=s_2\partial_{s_3}-s_3\partial_{s_2}.\] 
{  Structure relations:}
\[[L_1,X]=2L_2,\
 [L_2,X]= -X^2-2L_1+H-a,\ 
[L_1,L_2]=-(
 L_1X+XL_1)-(\frac12+2a)X,\]
\[ \frac13\left(X^2L_1+XL_1X+L_1X^2\right)+L_1^2+L_2^2-HL_1+(a+\frac{11}{12})X^2-\frac16
 H
+(a-\frac23)L_1-\frac{5a}{6}=0.\]
\item {$E3$ (Harmonic Oscillator)}
\[ H=\partial_x^2+\partial_y^2-\omega^2(x^2+y^2)\]
{ Basis symmetries:} 
\[L_1=\partial_x^2-\omega^2 x^2,\  L_3= \partial_{xy}-\omega^2 x y,\  X=x\partial_y-y\partial_x.\]
 Also we set $L_2=\partial_y^2-\omega^2 y^2=H-L_1$.

{ Structure equations:}                    
\[ [L_1,X] =2L_3,\  [L_3,X]=H-2L_1,\  [L_1,L_3]=2\omega^2 X,\]
\[L_1^2+L_3^2-L_1H-\omega^2X^2+\omega^2=0\]
\item {$E4$}
 \[ H =\partial_x^2+\partial_y^2+ a(x+iy)\]
{Basis Symmetries:} (with  $M=x\partial_y-y\partial_x$)
\[L_1=\partial_x^2+a x,\ L_2=\frac{i}{2}\lbrace M,X\rbrace -\frac{a}{4}(x+iy)^2,\  X=\partial_x+i\partial_y\]
{ Structure equations:} 
 \[ [L_1,X]=a,\  [L_2,X]=X^2,\
 [L_1,L_2]=X^3+HX-\left \lbrace L_1,X\right \rbrace,\]
 \[X^4-2 \left\lbrace L_1,X^2\right\rbrace +2HX^2+H^2+4a L_2=0\]
\item{$E5$}
 \[ H=\partial_x^2+\partial_y^2 +ax\]
{Basis  symmetries:} (where $M=x\partial_y-y\partial_x$)
\[ L_1=\partial_{xy}+\frac1{2} ay,\  L_2=\frac1{2}\lbrace M,X \rbrace -\frac1{4}ay^2,\  X=\partial_y\]
{ Structure equations:}
\[[L_1,L_2]=2X^3-HX,\ [L_1,X]=-\frac{a}{2},\  [L_2,X]=L_1,\]
\[X^4-HX^2+L_1^2+a L_2=0\]
\item {$E6$}
\[ H=\partial_x^2+\partial_y^2 + \frac{a}{x^2}\]
{ Basis  symmetries:} ($M=x\partial_y-y\partial_x$)
\[L_1=\frac12 \lbrace M,\partial_{x} \rbrace -\frac{ay}{x^2},\  L_2=M^2+\frac{ay^2}{x^2},\  X=\partial_{y}\]
{ Structure equations:}  
\[[L_1,L_2]=\{X,L_2\}+(2a+\frac12)X,\  [L_1,X]=H-X^2,\  [L_2,X]=2L_1,\]
\[L_1^2+\frac14\{L_2,X^2\}+\frac12 XL_2X-L_2H+(a+\frac34)X^2=0\]
\item {$E14$}
 \[ H=\partial_x^2+\partial_y^2  +\frac{b}{\overline{z}^2}\] 
{ Basis symmetries:}   (with $M=x\partial_y-y\partial_x$, $z=x+iy,\overline{z}=x-iy,$)
\[X=\partial_x-i\partial_y,\ 
L_1=\frac{i}{2}\lbrace M,X \rbrace+\frac{b}{\overline{z}},\  L_2=M^2+\frac{bz}{\overline{z}}\]
{ Structure equations:} 
\[ [L_1,L_2]=-\lbrace X,L_2 \rbrace -\frac{1}{2}X,\
[X, L_1]=-X^2,\  [X, L_2]=2L_1,\]
\[L_1^2+XL_2X-b H-\frac{1}{4}X^2=0\]
 \end{enumerate}

\section{Contractions of superintegrable systems}
  Suppose we have a nondegenerate quantum superintegrable system with generators $H,L_1,L_2$, $R=[L_1,L_2]$ and the usual structure equations, defining a quadratic algebra $Q$.
If we make a change of basis to new generators ${\tilde H},{\tilde L_1}, {\tilde L_2}$ and parameters ${\tilde a_1},{\tilde a_2}, {\tilde a_3}$ such that 
\[
\left(\begin{array}{c}
{\tilde L_1}\\
{\tilde L_2} \\
{\tilde H}
\end{array}\right)
=\left(\begin{array}{ccc}
A_{1,1} & A_{1,2}&A_{1,3} \\
A_{2,1}&A_{2,2} &A_{2,3}  \\
0 &0 &A_{3,3}
\end{array}\right)
\left(\begin{array}{c}
L_1\\
L_2 \\
H
\end{array}\right)+
\left(\begin{array}{ccc}
B_{1,1} & B_{1,2}&B_{1,3} \\
B_{2,1}&B_{2,2} &B_{2,3}  \\
B_{3,1} &B_{3,2} &B_{3,3}
\end{array}\right)
\left(\begin{array}{c}
a_1\\
a_2 \\
a_3
\end{array}\right),\]
\[ \left(\begin{array}{c}
{\tilde a_1}\\
{\tilde a_2} \\
{\tilde a_3}
\end{array}\right)
=\left(\begin{array}{ccc}
C_{1,1} & C_{1,2}&C_{1,3} \\
C_{2,1}&C_{2,2} &C_{2,3}  \\
C_{3,1} &C_{3,2} &C_{3,3}
\end{array}\right)
\left(\begin{array}{c}
a_1\\
a_2 \\
a_3
\end{array}\right)
\]
for some $3\times 3$ constant matrices $A=(A_{i,j}),B,C$ such that $\det A \cdot \det C\ne 0$,  we will have the same 
system with new structure equations  of the same form  for ${\tilde R}=[{\tilde L_1},{\tilde L_2}]$, $[{\tilde L_j},
{\tilde R}]$,  ${\tilde R}^2$,  but with transformed structure constants.
\begin{itemize}
\item
{ Choose a continuous 1-parameter family of basis transformation matrices} $A(\epsilon),B(\epsilon),C(\epsilon$, $0<\epsilon\le 1$ such 
that $A(1)=C(1)$ is the identity matrix, $B(1)=0$ and  $\det A(\epsilon)\ne 0$, $\det C(\epsilon)\ne 0$. 
\item Now 
{ suppose as $\epsilon\to 0$ the basis change becomes singular}, (i.e., the limits of $A,B,C$ either do not exist or, if they exist
 do not satisfy $\det A(0)\det C(0)\ne 0$) { but the structure equations} involving $A(\epsilon),B(\epsilon),C(\epsilon)$, {go to a limit,
  defining a new quadratic algebra $Q'$}.
\item We call $Q'$ a { contraction} of $Q$ in analogy with Lie algebra contractions.
\end{itemize}
{ There is a similar definition of a contraction of a degenerate superintegrable system.}
Further,  we say that the 2D system without potential,
$
 H_0=\Delta_2$,
and with 
 3  algebraically  independent second-order 
symmetries is a 2nd
  order free triplet. The possible spaces admitting free triplets are just those classified by Koenigs.
Note that every nondegenerate or degenerate superintegrable system defines a free triplet, simply by setting the parameters $a_j=0$ in the potential.
 Similarly, this free triplet defines a { free quadratic algebra}, i.e., a quadratic algebra with all $a_j=0$.
 {  In general, a free triplet cannot be obtained as a restriction of a superintegrable system and its associated algebra does not close to a free quadratic algebra.} All of these definitions extend easily to classical superintegrable systems.

We have the following {closure theorems:
 \begin{theorem} Closure Theorem: A free triplet (classical or quantum)  extends to a superintegrable system if and only if it generates a free quadratic algebra.
  \end{theorem}
\begin{theorem} A superintegrable system, degenerate or nondegenerate, is uniquely determined by its free quadratic algebra.
 \end{theorem}
Proofs of these results will appear in  \cite{KMSH}. The main ideas are as follows.
Suppose we have a classical free triplet with basis
\[ {\cal L}_{(s)}=\sum_{i,j=1}^2a^{ij}_{(s)} p_ip_j\quad a^{ij}_{(s)}=a^{ji}_{(s)},\ s=1,2,3,\ {\cal L}_{(3)}={\cal H}_0=\frac{p_1^2+p_2^2}{\lambda(x,y)},\] 
 that determines a free nondegenerate quadratic algebra, hence  a free nondegenerate superintegrable system. 
From the free system alone we can compute the functions $A^{ij},B^{ij}$, expressed in terms of the Cartesian-like coordinates $(x,y)$, 
that determine the system of equations for an additive potential 
\be\label{nondegpot1} \ba{lllll}
V_{22}&=&V_{11}&+&A^{22}V_1+B^{22}V_2,\\
V_{12}&=& &&A^{12}V_1+B^{12}V_2,\ea
\ee
These equations always admit a constant potential for a solution, but they will admit a full 4-dimensional vector space of solutions $V$ if and only if the
 integrability conditions for  (\ref{nondegpot1}) are identically satisfied. In \cite{KMSH} we show that the integrability conditions hold if and only if the free 
system generates a quadratic algebra. This is an algebraic solution  for an analytic problem.  Further, if a potential function satisfies 
 (\ref{nondegpot1}) then it is guaranteed that the Bertrand-Darboux integrability conditions for equations
 \[ W^{(s)}_i=\lambda\sum_{j=1}^2 a^{ij}_{(s)}V_j,\quad i,s=1,2,\] 
 hold and we can compute the solutions $ W^{(s)}$, $W^{(3)}=V$, unique up to  additive constants, such that the constants of the motion
${\cal L}_{(s)} =\sum
a^{ij}_{(s)}p_ip_j+W^{(s)}$ define a nondegenerate superintegrable system. This system is guaranteed to determine  a nondegenerate quadratic algebra with 
potential whose highest 
order (potential-free) terms agree with the free quadratic algebra. The functions $A^{ij},B^{ij}$ are defined independent of the basis chosen 
for the free triplet although, of course, 
they do depend upon the particular coordinates chosen.

Similarly, there is an associated  2nd
  order quantum free triplet
\[  L_s=\frac{1}{\lambda}\sum_{i,j=1}^2\partial_{i}(\lambda a^{ij}_{(s)}\partial_j)
,\ s=1,2,3,\  L_3=H_0=\frac{1}{\lambda({\bf x})}(\partial_{11}+\partial_{22}),\]
that defines  a free nondegenerate quantum quadratic algebra with potential. The functions $W^{(s)}$ are the same as before.

There is an analogous construction of degenerate superintegrable systems with potential from  free triplets that generate a free  quadratic algebras, 
but are such that one generator say, ${\cal L}_1={\cal X}^2$, is a perfect square. 

\section{Lie algebra contractions}
The contractions of the Lie algebras $e(2,{\bf C})$ and $o(3,{\bf C})$} have long since been classified, e.g.  \cite{WW}. There are 7 nontrivial contractions of 
 $e(2,{\bf C})$ and 4 of $o(3,{\bf C})$. However, 2 of the contractions of $e(2,{\bf C})$ take it to an abelian Lie algebra so are not of interest to us.

\medskip\noindent 
{\bf Wigner-Inonu contractions of $e(2,{\bf C})$:}
\begin{enumerate}
\item $ \{{\cal J}',p_1',p_2'\}=\{ {\cal J},\ \epsilon p_1,\ \epsilon p_2\}: \  e(2,{\bf C}),$\hfill  \break
$ {\rm coordinate\ implementation}\ x'=\frac{x}{\epsilon},y'=\frac{y}{\epsilon},$
\item $ \{{\cal J}',p_1',p_2'\}=\{\epsilon {\cal J},\ p_1,\ \epsilon p_2\}\ : \ {\rm Heisenberg\ algebra},$\hfill\break
${\rm coordinate\ implementation}\ x'=x,y'=\frac{y}{\epsilon}, {\cal J}'=x'p_2'$,
\item $ \{  {\cal J}',\  p_1'+ip_2',\ p_1'-ip_2'\}=\{ \epsilon {\cal J},\ \epsilon( p_1+ip_2),\ p_1-ip_2\}\ :\ {\rm abelian\ algebra},$
\item $ \{{\cal J}',p_1',p_2'\}=\{\epsilon {\cal J},\ p_1,\ p_2\}\ : \ {\rm abelian\ algebra},$
\item $\{  {\cal J}',\  p_1'+ip_2',\ p_1'-ip_2'\}=\{{\cal J},\ \epsilon(p_1+ip_2),\ p_1-ip_2\}\ :\ e(2,{\bf C}),$\hfill\break
${\rm coordinate\ implementation}\ x'+iy'=x+iy,x'-iy'=\frac{x-iy}{\epsilon},$
\end{enumerate}

\medskip\noindent 
{\bf  The other natural contractions of $e(2,{\bf C})$:}
\begin{enumerate}\setcounter{enumi}{5}
\item 
 $ \{{\cal J}',p_1',p_2'\}=\{ {\cal J}+\frac{p_1}{\epsilon},\ p_1,\ p_2\}:\   e(2,{\bf C}),$\hfill\break
${\rm coordinate\ implementation}\ x'=x,y'=y-\frac{1}{\epsilon},$
 \item $\{{\cal J}',p_1',p_2'\}=\{ {\cal J}+\frac{p_1+ip_2}{\epsilon},\ p_1,\  p_2\}:\  e(2,{\bf C}),$\hfill\break
${\rm coordinate\ implementation}\ x'=x+\frac{i}{\epsilon},y'=y-\frac{1}{\epsilon}.$
\end{enumerate}

We use the classical realization for $o(3,{\bf C})$ acting on the 2-sphere, with basis ${\cal J}_1=s_2p_3-s_3p_2,\ {\cal J}_2=s_3p_1-s_1p_3,\ {\cal J}_3=s_1p_2-s_2p_1$, commutation relations
\[ \{{\cal J}_2,{\cal J}_1\}={\cal J}_3,\quad \{{\cal J}_3,{\cal J}_2\}={\cal J}_1,\quad \{{\cal J}_1,{\cal J}_3\}={\cal J}_2,\]
and Hamiltonian ${\cal H}={\cal J}_1^2+{\cal J}_2^2+{\cal J}_3^2$. Here $s_1^2+s_2^2+s_3^2=1$ and restriction to the sphere gives
$s_1p_1+s_2p_2+s_3p_3=0$.

\bigskip\noindent 
{\bf Wigner-Inonu contractions of $o(3,{\bf C})$:}
\begin{enumerate}
\item $\{{\cal J}_1',{\cal J}_2',{\cal J}_3'\}=\{ \epsilon {\cal J}_1,\ \epsilon {\cal J}_2,\  {\cal J}_3\}:\  e(2,{\bf C}),$\hfill\break
${\rm coordinate\ implementation}\ x={s_1}/{\epsilon},y={s_2}/{\epsilon}, s_3\approx 1, {\cal J}={\cal J}_3,$
\item $\{ {\cal J}_1'+i{\cal J}_2',\  {\cal J}_1'-i{\cal J}_2',\  {\cal J}_3'\}=\{ {\cal J}_1+i{\cal J}_2,\  \epsilon({\cal J}_1-i{\cal J}_2),\ \epsilon {\cal J}_3\}:\ {\rm Heisenberg\ algebra},$ \hfill\break
${\rm coordinate\ implementation}\ s_1=\frac{\cos\phi}{\cosh\psi},\  s_2=\frac{\sin\phi}{\cosh\psi},\  s_3=\frac{\sinh\psi}{\cosh\psi},$\hfill\break
${\rm we\ set}\ \phi=\epsilon\theta-i\ln \sqrt{\epsilon},\ \psi=\xi\sqrt{\epsilon},\ {\rm to\ get\ }$\hfill\break
$  {\cal J}_3'=p_\theta,\ {\cal J}_1'+i{\cal J}_2'=-i(\xi p_\theta+p_\xi),\ {\cal J}_1'-i{\cal J}_2'=-\xi p_\theta+p_\xi,$
\item $ \{ {\cal J}_1'+i{\cal J}_2',\  {\cal J}_1'-i{\cal J}_2',\  {\cal J}_3'\}=\{ {\cal J}_1+i{\cal J}_2,\  \epsilon({\cal J}_1-i{\cal J}_2),\  {\cal J}_3\}:\  e(2,{\bf C}),$\hfill\break
${\rm coordinate\ implementation}\ s_1+is_2=\epsilon z,\ s_1-is_2= {\bar z},\ s_3\approx 1,$\hfill\break
$ {\rm Using}\ zp_z+{\bar z}p_{\bar z}+s_3 p_{s_3}=0,\
 {\rm we \ get}\ {\cal J}_3'=i(zp_z-{\bar z}p_{\bar z}),$\hfill\break
${\cal J}_1'+i{\cal J}_2' =2ip_{\bar z},\ {\cal J}_1'-i{\cal J}_2'=-2ip_z,$
 \end{enumerate}

\medskip\noindent 
{\bf  The other natural contraction of $o(3,{\bf C})$:}
\begin{enumerate}\setcounter{enumi}{4}
\item $  \{{\cal J}'_1+i{\cal J}_2',{\cal J}_1'-i{\cal J}_2',{\cal J}_3'\}=\{ \epsilon({\cal J}_1+i{\cal J}_2),\frac{{\cal J}_1-i{\cal J}_2}{\epsilon},{\cal J}_3\}:\   o(3,{\bf C}),$\hfill\break
${\rm coordinate\ implementation}\ 
 s_1'=\frac{\epsilon+\epsilon^{-1}}{2}s_1+i\frac{\epsilon-\epsilon^{-1}}{2}s_2,$\hfill\break
$ s_2'=-i\frac{\epsilon-\epsilon^{-1}}{2}s_1+\frac{\epsilon+\epsilon^{-1}}{2}s_2,\  s_3'=s_3.$
\end{enumerate}

Note that once we choose a basis for a Lie algebra $A$, the structure of its enveloping algebra is uniquely determined by the structure constants. 
All structure relations in the enveloping
algebra are continuous functions of the structure constants. Thus a contraction of one Lie algebra $A$ to another, $B$  induces a similar contraction of the 
corresponding  enveloping algebras of $A$ and $B$. In the case of $e(2,{\bf C})$ and $o(3,{\bf C})$, free quadratic algebras constructed in the enveloping algebras will contract 
to free quadratic algebras generated by the target Lie algebras, \cite{KMSH}
We illustrate the process with several examples.
In the following examples we work out all of the induced contractions for the systems  $\tilde E1$, $\tilde S9$, $\tilde S3$ and $\tilde E3$ to illustrate the
 contraction procedure for each of these Lie 
algebras and for both  nondegenerate and degenerate systems.

\begin{enumerate}
 \item {\bf  ${\tilde E1}\to {\tilde E8}$:} Use $\{  {\cal J}',\  p_1'+ip_2',\ p_1'-ip_2'\}=\{  {\cal J},\  \epsilon(p_1+ip_2),\ p_1-ip_2\}$.
\begin{eqnarray*} {\cal L}_1&=&{\cal J}^2=({\cal J}')^2={\cal L}_1'\\
 {\cal L}_2&=&p_1^2=\frac14\left((p_1+ip_2)+(p_1-ip_2)\right)^2=\frac14\left(\frac{(p_1'+ip_2')}{\epsilon}+(p_1'-ip_2')\right)^2\\
&\approx& \frac{(p_1'+ip_2')^2}{4\epsilon^2}=\frac{{\cal L}_2'}{4\epsilon^2}\\
{\cal H}&=&(p_1+ip_2)(p_1-ip_2)=\frac{(p_1'+ip'_2)(p_1'-ip_2')}{\epsilon}=\frac{ {\cal H}'}{\epsilon}.
  \end{eqnarray*}

 \item  $\tilde E1\to \tilde E2$: Use $ \{{\cal J}',p_1',p_2'\}=\{ {\cal J}+\frac{p_2}{\epsilon},\ p_1,\ p_2\}$.
\begin{eqnarray*} {\cal L}_1&=&{\cal J}^2=({\cal J}'-\frac{p_2'}{\epsilon})^2=({\cal J}'^2)-2\frac{p_2'{\cal J}'}{\epsilon}+\frac{(p_2')^2}{\epsilon^2}
\approx -2\frac{{\cal L}_2'}{\epsilon}+\frac{{\cal L}_1'}{\epsilon^2}\\
 {\cal L}_2&=&p_1^2={(p_1')}^2={\cal H}'-{\cal L}_1',\quad {\cal L}_2'={\cal J}'p_2'\\
{\cal H}&=&p_1^2+p_2^2={(p_1')}^2+{(p_2')}^2= {\cal H}'.
  \end{eqnarray*}

 \item  $\tilde E1\to \tilde E3'$: Use $ \{{\cal J}',p_1',p_2'\}=\{ {\cal J}+\frac{p_1+p_2}{\epsilon},\ p_1,\  p_2\}$.
\begin{eqnarray*} {\cal L}_1&=&{\cal J}^2= ({\cal J}'-\frac{p_1'+p_2'}{\epsilon})^2={{\cal J}'}^2-2\frac{{\cal J}'(p'_1+p'_2)}{\epsilon}+\frac{{p_1'}^2+2p_1'p_2'+{p_2'}^2}{\epsilon^2}\\
&\approx &\frac{{\cal L}_2'}{\epsilon^2} +\frac{{\cal H}'}{\epsilon^2}\\
 {\cal L}_2&=&p_1^2={\cal H}'-{\cal L}_1',\quad {\cal L}_2'=2p_1'p_2'\\
{\cal H}&=&p_1^2+p_2^2= {\cal H}'.
  \end{eqnarray*}

\item $\tilde E1\to \tilde E3'$ (alternate version). Use $\{{\cal J}',p_1',p_2'\}=\{{\cal J}+(p_1+ip_2)/\epsilon,p_1,p_2\}$.
\begin{eqnarray*} {\cal L}_1'&=&p_1p_2=\frac{\epsilon^2 {\cal L}_1+{\cal H}-2{\cal L}_2}{2i} \\
 {\cal L}_2'&=&p_1^2={\cal L}_2\\
{\cal H}'&=&p_1^2+p_2^2= {\cal H}.
  \end{eqnarray*}
\item $\tilde E1\to \tilde  E1$.  Use $\{{\cal J}',p_1',p_2'\}=\{{\cal J},\epsilon p_1,\epsilon p_2\}$.
\begin{eqnarray*} {\cal L}_1'&=&{\cal J}^2={\cal L}_1\\
 {\cal L}_2'&=&\epsilon^2 p_1^2=\epsilon^2 {\cal L}_2\\
{\cal H}'&=&\epsilon^2(p_1^2+p_2^2)= \epsilon^ 2 {\cal H}.
  \end{eqnarray*}
\item $\tilde E_1\to \ {\rm Heisenberg}$. Use $\{{\cal J}',p_1',p_2'\}=\{\epsilon {\cal J},p_1,\epsilon p_2\}$. 
\begin{eqnarray*} {\cal L}_1&=&{\cal J}^2=\frac{{{\cal J}'}^2}{\epsilon^2}\\
 {\cal L}_2&=& p_1^2=p{_1'}^2\\
{\cal H}&=&p_1^2+p_2^2={p_1'}^2+\frac{{p_2'}^2}{\epsilon^2}={\cal L}_2+ \frac{{p_2'}^2}{\epsilon^2},
  \end{eqnarray*}
so,
\begin{eqnarray*} {\cal L}_1'&=&\epsilon^2 {\cal L}_1={{\cal J}'}^2={x'}^2{p_2'}^2\\
 {\cal L}_2'&=& {\cal L}_2={p_1'}^2\\
{\cal H}'&=&\epsilon^2({\cal H}-{\cal L}_2)={p_2'}^2.
  \end{eqnarray*}
Structure relations:
\[ {\cal R}=\{{\cal L}_1',{\cal L}_2'\},\  {\cal R}^2=4{p_1'}^2{p_2'}^4=4{\cal L}_1'{{\cal H}'}^2.\]

\item $\tilde S9\to \tilde E1$: Use $\{{\cal J}_1',{\cal J}_2',{\cal J}_3'\}=\{ \epsilon {\cal J}_1,\ \epsilon {\cal J}_2,\  {\cal J}_3\}$.
\begin{eqnarray*}
{\cal L}_1&=&{\cal J}_3^2={\cal J}^2={\cal L}_1'\\
{\cal L}_2&=&{\cal J}_1^2\approx \frac{{p_1'}^2}{\epsilon^2}=\frac{{{\cal L}_2'}^2}{\epsilon^2} \\
{\cal H}&=& {\cal J}_1^2+{\cal J}_2^2+{\cal J}_3^2\approx \frac{{p_1'}^2+{p_2'}^2}{\epsilon^2}=\frac{{\cal H}'}{\epsilon^2}
 \end{eqnarray*}

\item $\tilde S9\to \tilde S2$: Use $\{{\cal J}'_1+i{\cal J}_2',{\cal J}_1'-i{\cal J}_2',{\cal J}_3'\}=\{ \epsilon({\cal J}_1+i{\cal J}_2),\frac{{\cal J}_1-i{\cal J}_2}{\epsilon},{\cal J}_3\}$.
\begin{eqnarray*}
 {\cal L}_1&=& {\cal J}_3^2={{\cal J}_3'}^2={\cal L}_2'\\
{\cal L}_2&=& {\cal J}_1^2=\frac14\left( ({\cal J}_1+i{\cal J}_2)+({\cal J}_1-i{\cal J}_2)\right)^2=\frac14\left( \frac{{\cal J}_1'+i{\cal J}_2'}{\epsilon}+\epsilon({\cal J}_1'-i{\cal J}_2')\right)^2\approx \frac14
 \frac{{\cal L}_1'}{\epsilon^2}\\
{\cal H}&=& ({\cal J}_1+i{\cal J}_2)({\cal J}_1-i{\cal J}_2)+{\cal J}_3^2 ={{\cal J}_1'}^2+{{\cal J}_2'}^2+{{\cal J}_3'}^2={\cal H}',\quad {\cal L}_1'=({\cal J}_1'+i{\cal J}_2')^2,
\end{eqnarray*}
so the change of basis
\begin{eqnarray*}
 {\cal L}_1'&=& 4\epsilon^2 {\cal L}_2,\\
{\cal L}_2'&=& {\cal L}_1,\\
{\cal H}'&=&{\cal H}.
\end{eqnarray*}
determines the contraction to $\tilde S2$ in the limit as $\epsilon\to 0$.

\item $\tilde S9\to \tilde E8$: Use $\{ {\cal J}_1'+i{\cal J}_2',\  {\cal J}_1'-i{\cal J}_2',\  {\cal J}_3'\}=\{ {\cal J}_1+i{\cal J}_2,\  \epsilon({\cal J}_1-i{\cal J}_2),\  {\cal J}_3\}$,
with coordinate implementation $ s_1+is_2=\epsilon z$, $ s_1-is_2= {\bar z},\ s_3\approx 1$,
so ${\cal J}_3'=i(zp_z-{\bar z}p_{\bar z})$,${\cal J}_1'+i{\cal J}_2' =2ip_{\bar z}$, ${\cal J}_1'-i{\cal J}_2'=-2ip_z$.
\begin{eqnarray*}
 {\cal L}_1&=& {\cal J}_3^2={{\cal J}_3'}^2={\cal L}_1'\\
{\cal L}_2&=& {\cal J}_1^2=\frac14\left( ({\cal J}_1+i{\cal J}_2)+({\cal J}_1-i{\cal J}_2)\right)^2=\frac14\left( ({\cal J}_1'+i{\cal J}_2')+\frac{{\cal J}_1'-i{\cal J}_2'}{\epsilon}\right)^2\approx \frac14
 \frac{{\cal L}_2'}{\epsilon^2}\\
{\cal H}&=& ({\cal J}_1+i{\cal J}_2)({\cal J}_1-i{\cal J}_2)+{\cal J}_3^2 =\frac{({\cal J}_1'+i{\cal J}_2')({\cal J}_1'-i{\cal J}_2')}{\epsilon}+{{\cal J}_3'}^2=\frac{{\cal H}'}{\epsilon}+{\cal L}_1',
\end{eqnarray*}
where ${\cal L}_2'=({\cal J}_1'-i{\cal J}_2')^2$,
so the change of basis
\begin{eqnarray*}
 {\cal L}_1'&=&  {\cal L}_1,\\
{\cal L}_2'&=& 4\epsilon^2{\cal L}_2,\\
{\cal H}'&=&\epsilon {\cal H},
\end{eqnarray*}
determines the contraction to $\tilde E8$ in the limit as $\epsilon\to 0$.
\item $\tilde S9\to \ {\rm Heisenberg\ algebra}$: Use $\{ {\cal J}_1'+i{\cal J}_2',\  {\cal J}_1'-i{\cal J}_2',\  {\cal J}_3'\}=\{ {\cal J}_1+i{\cal J}_2,\  \epsilon({\cal J}_1-i{\cal J}_2),\ \epsilon {\cal J}_3\}$.
with  coordinate implementation $s_1=\frac{\cos\phi}{\cosh\psi}$, $s_2=\frac{\sin\phi}{\cosh\psi}$, $s_3=\frac{\sinh\psi}{\cosh\psi}$,
and substitutions  $\phi=\epsilon\theta-i\ln \sqrt{\epsilon}$, $\psi=\xi\sqrt{\epsilon}$,
 to\ get  ${\cal J}_3'=p_\theta$, ${\cal J}_1'+i{\cal J}_2'=-i(\xi p_\theta+p_\xi)$, ${\cal J}_1'-i{\cal J}_2'=-\xi p_\theta+p_\xi$.
\begin{eqnarray*}
 {\cal L}_1&=& {\cal J}_3^2=\frac{{{\cal J}_3'}^2}{\epsilon^2}=\frac{{\cal L}_1'}{\epsilon^2}\\
{\cal L}_2&=& {\cal J}_1^2=\frac14\left( ({\cal J}_1+i{\cal J}_2)+({\cal J}_1-i{\cal J}_2)\right)^2=\frac14\left( ({\cal J}_1'+i{\cal J}_2')+\frac{{\cal J}_1'-i{\cal J}_2'}{\epsilon}\right)^2\approx \frac14
 \frac{{\cal L}_2'}{\epsilon^2}\\
{\cal H}&=& ({\cal J}_1+i{\cal J}_2)({\cal J}_1-i{\cal J}_2)+{\cal J}_3^2 =\frac{{{\cal J}_1'}^2+{{\cal J}_2'}^2}{\epsilon}+\frac{{{\cal J}_3'}^2}{\epsilon^2},
\end{eqnarray*}
where ${\cal L}_1'=({\cal J}_1'-i{\cal J}_2')^2$, ${\cal L}_2'={{\cal J}_1'}^2+{{\cal J}_2'}^2$.
so the change of basis
\begin{eqnarray*}
 {\cal H}'&=&  \epsilon^2 {\cal H},\\
{\cal L}_1'&=& 4\epsilon^2{\cal L}_2,\\
{\cal L}_2'&=&\epsilon ({\cal H}-{\cal L}_1),
\end{eqnarray*}
${{\cal R}'}^2=-16{\cal H}'{{\cal L}'_1}^2$, determines the contraction.

\item $\tilde S3\to \tilde E3$. Use $\{{\cal J}_1',{\cal J}_2',{\cal J}_3'\}=\{\epsilon {\cal J}_1,\epsilon {\cal J}_2, {\cal J}_3\}$. 
\begin{eqnarray*}
{\cal X}'&=& {\cal X},\\
 {\cal L}_1'&=&  \epsilon^2 {\cal L}_1,\\
{\cal L}_2'&=& \epsilon^2{\cal L}_2,\\
{\cal H}'&=&\epsilon^2 {\cal H}.
\end{eqnarray*}
\item $\tilde S3\to \tilde E3$ (alternate contraction). Use $\{{\cal J}_1'+i{\cal J}_2',{\cal J}_1'-i{\cal J}_2',{\cal J}_3'\}=\{ {\cal J}_1+i{\cal J}_2,\epsilon({\cal J}_1-i {\cal J}_2), {\cal J}_3\}$. 
\begin{eqnarray*}
{\cal X}'&=& {\cal X}=i(zp_z-{\bar z}p_{\bar z}),\\
 {\cal L}_1'&=& p_{\bar z}^2=-\frac{1}{2}({\cal L}_1+i{\cal L}_2)+\frac14 {\cal H}-\frac14 {\cal X}^2,\\
{\cal L}_2'&=& p_z^2=-i\epsilon^2{\cal L}_2,\\
{\cal H}'&=&4p_zp_{\bar z}=\epsilon  {\cal H}.
\end{eqnarray*}
\item $\tilde S3\to \tilde S3$. Use $\{{\cal J}_1'+i{\cal J}_2',{\cal J}_1'-i{\cal J}_2',{\cal J}_3'\}=\{ \epsilon({\cal J}_1+i{\cal J}_2),({\cal J}_1-i {\cal J}_2)/\epsilon, {\cal J}_3\}$. 
\begin{eqnarray*}
{\cal X}'&=& {\cal X}={\cal J}_3,\\
 {\cal L}_1'&=& ({\cal J}_1'+i{\cal J}_2')^2=4i\epsilon^2{\cal L}_2,\\
{\cal L}_2'&=& ({\cal J}_1'-i{\cal J}_2')^2=\frac{2}{\epsilon^2}({\cal L}_1-i{\cal L}_2-\frac12 {\cal H}+\frac12 {\cal X}^2),\\
{\cal H}'&=&{{\cal J}_1'}^2+{{\cal J}_2'}^2+{{\cal J}_3'}^2=  {\cal H}.
\end{eqnarray*}
\item $\tilde S3\to \ {\rm Heisenberg}$. Use $\{{\cal J}_1'+i{\cal J}_2',{\cal J}_1'-i{\cal J}_2',{\cal J}_3'\}=\{{\cal J}_1+i{\cal J}_2,\epsilon({\cal J}_1-i{\cal J}_2),\epsilon {\cal J}_3\}$.
\begin{eqnarray*}
{\cal X}'&=& {\cal J}_3'=\epsilon {\cal X},\\
 {\cal L}_1'&=& ({\cal J}_1'+i{\cal J}_2')^2={\cal L}_1,\\
{\cal L}_2'&=& ({\cal J}_1'-i{\cal J}_2')^2=\epsilon^2 {\cal L}_2,\\
{\cal H}'&=&{{\cal J}_3'}^2=  \epsilon^2 {\cal H}.
\end{eqnarray*}
The structure relation is ${\cal H}'-{{\cal X}'}^2=0$.
\item $\tilde E3\to \tilde E3$. Use $\{{\cal J}',p_1',p_2'\}=\{ {\cal J},\epsilon p_1,\epsilon p_2\}$. 
\begin{eqnarray*}
{\cal X}'&=& {\cal J}'={\cal X},\\
 {\cal L}_1'&=& {p_2'}^2=\epsilon^2{\cal L}_1,\\
{\cal L}_2'&=& p_1'p_2'=\epsilon^2 {\cal L}_2,\\
{\cal H}'&=&{p_1'}^2+{p_2'}^2=\epsilon^2  {\cal H}.
\end{eqnarray*}
\item $\tilde E3\to \tilde E3$ (alternate form contraction). Use $\{{\cal J}',p_1'+ip_2',p_1'-ip_2'\}=\{ {\cal J},\epsilon (p_1+ip_2),p_1-i p_2\}$. 
\begin{eqnarray*}
{\cal X}'&=& {\cal J}'={\cal X},\\
 {\cal L}_1'&=& (p_1'-i{p_2'})^2=2i({\cal L}_1-{\cal L}_2)-{\cal H},\\
{\cal L}_2'&=&(p_1'+i{p_2'})^2=4i\epsilon^2 {\cal L}_2,\\
{\cal H}'&=&{p_1'}^2+{p_2'}^2=\epsilon  {\cal H}.
\end{eqnarray*}
\item $\tilde E3\to \tilde E5$.  Use $\{{\cal J}',p_1',p_2'\}=\{ {\cal J}+\frac{p_1}{\epsilon},p_1, p_2\}$. 
 Take the 1st order basis for $\tilde E3$ as ${\cal X}={\cal J}$, with 2nd order basis ${\cal X}^2,{\cal L}_1=p_1^2,{\cal L}_2=p_1p_2,{\cal H}=p_1^2+p_2^2$.
 \begin{eqnarray*}
{\cal X}'&=& p_1'=\epsilon {\cal X},\\
 {\cal L}_1'&=&{p_1'}^2={\cal L}_1,\\
{\cal L}_2'&=&p_1'p_2'={\cal L}_2,\\
{\cal H}'&=&{p_1'}^2+{p_2'}^2= {\cal H}.
\end{eqnarray*}
However,  ${\cal L}_1'={{\cal X}'}^2$ 
so the space of 2nd order symmetries would appear to have dimension only 3. The missing 2nd order symmetry is constructed from ${\cal X}^2$ and ${\cal L}_1$: 
\[{\cal L}_3'={\cal J}'p_1'=-\frac{\epsilon}{2}({\cal X}^2-\frac{{\cal L}_1}{\epsilon^2}).\]
\item $\tilde E3\to \tilde E4$.  Use $\{{\cal J}',p_1',p_2'\}=\{ {\cal J}+\frac{p_1+ip_2}{\epsilon},p_1, p_2\}$. 
 Take the 1st order basis for $\tilde E3$ as ${\cal X}={\cal J}$, with 2nd order basis ${\cal X}^2,{\cal L}_1=p_2^2,{\cal L}_2=p_1p_2,{\cal H}=p_1^2+p_2^2$.
 \begin{eqnarray*}
{\cal X}'&=& p_1'+ip_2'=\epsilon {\cal X},\\
 {\cal L}_1'&=&{p_2'}^2={\cal L}_1,\\
{\cal L}_2'&=&p_1'p_2'={\cal L}_2,\\
{\cal H}'&=&{p_1'}^2+{p_2'}^2= {\cal H}.
\end{eqnarray*}
However,  ${\cal H}'-2{\cal L}_1'+2i{\cal L}_2'={{\cal X}'}^2$ 
so the space of 2nd order symmetries would appear to have dimension only 3. The missing 2nd order symmetry is constructed as
\[{\cal L}_3'={\cal J}'(p_1'+ip_2')=-\frac{\epsilon}{2}({\cal X}^2-\frac{{\cal H}+2i{\cal L}_2-2{\cal L}_1}{\epsilon^2}).\]
\item $\tilde E3\to \ {\rm Heisenberg}$.  Use $\{{\cal J}',p_1',p_2'\}=\{ \epsilon {\cal J},p_1, \epsilon p_2\}$. 
 Take the basis as  ${\cal X}={\cal J}$,and ${\cal X}^2,{\cal L}_1=p_1^2,{\cal L}_2=p_1p_2,{\cal H}=p_1^2+p_2^2$.
 \begin{eqnarray*}
{\cal X}'&=& {\cal J}'=\epsilon {\cal X},\\
 {\cal L}_1'&=&{p_1'}^2={\cal L}_1,\\
{\cal L}_2'&=&p_1'p_2'=\epsilon {\cal L}_2,\\
{\cal H}'&=&{p_2'}^2= \epsilon^2 {\cal H}.
\end{eqnarray*}
The functional relation is ${\cal L}'_1{\cal H}'-{{\cal L}'_2}^2=0$.
\end{enumerate}

Suppose we have a classical free triplet ${\cal H}^{(0)}, {\cal L}_1^{(0)}, {\cal L}_2^{(0)}$ that determines a nondegenerate quadratic algebra $Q^{(0)} $
and structure functions $A^{ij}({\bf x}), B^{ij}({\bf x})$ in some set of Cartesian-like coordinates $(x_1,x_2)$. Further, suppose this system contracts to another nondegenerate system
${{\cal H}'}^{(0)}, {{\cal L}'}_1^{(0)}, { {\cal L}'}_2^{(0)}$ with quadratic algebra ${Q'}^{(0)}$ via the mechanism described in the preceding sections. We show here that this contraction induces a contraction of the associated nondegenerate superintegrable system 
${\cal H}={\cal H}^{(0)}+V$, ${\cal L}_1={\cal L}_1^{(0)}+W^{(1)}$, 
 ${\cal L}_2={\cal L}_2^{(0)}+W^{(2)}$, $Q$ to
${\cal H}'={{\cal H}'}^{(0)}+V'$, ${\cal L}'_1={{\cal L}'}_1^{(0)}+{W^{(1)}}'$, 
 ${\cal L}'_2={{\cal L}'}_2^{(0)}+{W^{(2)}}'$, $Q'$. 
The point is that in  the contraction process the symmetries ${{\cal H}'}^{(0)}(\epsilon)$, 
${{\cal L}'}_1^{(0)}(\epsilon)$, 
${ {\cal L}'}_2^{(0)}(\epsilon)$  
remain continuous functions of $\epsilon$, linearly independent as quadratic forms, and
 $\lim_{\epsilon\to 0} {{\cal H}'}^{(0)}(\epsilon)={{\cal H}'}^{(0)}$, 
$\lim_{\epsilon\to 0} {{\cal L}'}^{(0)}_j(\epsilon)={{\cal L}'}^{(0)}_j$. 
Thus the associated functions $A^{ij}(\epsilon), B^{ij}(\epsilon)$ will also be continuous functions of $\epsilon$ and 
$\lim_{\epsilon\to 0}A^{ij}(\epsilon)={A'}^{ij}$, $\lim_{\epsilon\to 0}B^{ij}(\epsilon)={B'}^{ij}$. Similarly, the integrability conditions for the potential equations 
\be\label{nondegpot2} \ba{lllll}
 V^{(\epsilon)}_{22}&=& V^{(\epsilon)}_{11}&+&A^{22}(\epsilon) V^{(\epsilon)}_1+B^{22}(\epsilon) V^{(\epsilon)}_2,\\
 V^{(\epsilon)}_{12}&=& &&A^{12}(\epsilon) V^{(\epsilon)}_1+B^{12}(\epsilon) V^{(\epsilon)}_2,\ea
\ee
will hold for each $\epsilon$ and in the limit. This means that the 4-dimensional solution space for the potentials $V$ will deform continuously into the 4-dimensional solution space for the potentials $V'$. Thus the target space of solutions $V'$ is uniquely determined by the free quadratic algebra contraction. 

\begin{example}
We describe  the contraction of 
$S9$ to $E1$, including the potential terms. Recall for $S9$  in
  coordinates $x_1=\psi,x_2=\phi$  we have 
\[A^{12}=0,\quad A^{22}=\frac{3\cosh^2\psi-\sinh^2\psi}{\sinh\psi\cosh\psi},\quad B^{12}=2\frac{\sinh\psi}{\cosh\psi},\]
\[ B^{22}=-3\frac{(\cos^2\phi-\sin^2\phi)}{\sin\phi\cos\phi},\]
\be\label{S9pot}V=\frac{a_1\cosh^2\psi}{\cos^2\phi}+\frac{a_2\cosh^2\psi}{\sin^2\phi}+\frac{a_3\cosh^2\psi}{\sinh^2\psi}+a_4.\ee

For 
$E1$
and using polar coordinates $y_1=R,y_2=\phi'$ where $x=e^R\cos\phi',\ y=e^R\sin\phi'$, we have 
\[A'^{12}=0,\quad A'^{22}=-2,\quad B'^{12}=-2,\quad B'^{22}=-3\frac{(\cos^2\phi-\sin^2\phi)}{\sin\phi\cos\phi},\]
The general potential is 
\be\label{E1pot}V'=b_1e^{2R}+\frac{b_2e^{-2R}}{\cos^2\phi}+\frac{b_3e^{-2R}}{\sin^2\phi}+b_4.\ee

In terms of these coordinates the standard  contraction of the sphere to flat space 
is expressed as $\psi\approx \frac12\ln(\frac{1}{\epsilon})-R$, $\phi=\phi'$. In the limit as $\epsilon\to 0$ we have
\[A^{12}\to A'^{12}=0,\ A'^{22}\to 2=-A'^{22},\ B^{12}\to 2=-B'^{12},\]
\[ B^{22}=-3\frac{(\cos^2\phi-\sin^2\phi)}{\sin\phi\cos\phi}=B'^{22}.\]
The change in sign for $A^{22}$ and $B^{12}$ is due to the fact that $y_1$ corresponds to $-x_1$ whereas $y_2$ corresponds to $x_2$.
In the limit the 4 dimensional space of potentials (\ref{S9pot}) must go to the 4 dimensional vector space (\ref{E1pot}).
However the basis functions for the $S9$ potential,
\[ \frac{\cosh^2\psi}{\cos^2\phi},\ \frac{\cosh^2\psi}{\sin^2\phi},\ \frac{\cosh^2\psi}{\sinh^2\psi},\ 1\]
will not go to a new basis in the limit; 2 basis functions become unbounded and 2 go to a constant. There are many ways to choose an $\epsilon$ dependent basis so 
that the limit can be taken. One of the simplest choices of basis is
\[ V^{(1)}(\epsilon)=\frac{1}{4\epsilon}(\frac{\cosh^2\psi}{\sinh^2\psi}-1)\to e^{2R},\  \]
\[V^{(2)}(\epsilon)=\epsilon\frac{\cosh^2\psi}{\cos^2\phi}\to \frac{e^{-2R}}{\cos^2\phi},\ 
 V^{(3)}(\epsilon)=\epsilon\frac{\cosh^2\psi}{\sin^2\phi}\to \frac{e^{-2R}}{\sin^2\phi},\  V^{(4)}(\epsilon)=1\to 1.
\]
\end{example}

\section{Models of superintegrable systems}
\begin{itemize}\item
A { representation of a quadratic algebra $Q$} is a homomorphism of $Q$ into the associative  algebra of linear operators on some vector space:
 products go to products, commutators to commutators, etc.
 \item A { model $M$} is  a
 faithful  representation of $Q$ in which the vector space is a space of polynomials in one complex variable and the action is via   differential/difference 
operators acting on that space. 
 We  study classes of irreducible representations realized by these models. 
 \item Suppose a quadratic algebra $Q$ contracts to a  algebra $Q'$ via a 
continuous family of transformations indexed by  $\epsilon$. If we have a model $M$ of $Q$ we can 
try to ``save" this representation by passing through a continuous family of  models $M(\epsilon)$ of $Q(\epsilon)$ to
obtain a model $M'$  of $Q'$. 
\item There are three closely related limits tying one superintegrable system to another: 1) The pointwise coordinate limit of the source physical system  to the target  system. 2) The induced contraction of the source quadratic algebra to the target quadratic algebra. 3) The process of saving a representation of the target quadratic algebra by passing through a continuous family of models of representations of the source quadratic algebra, see Figure 1,
\item { As a byproduct of contractions to  systems from $S9$ for which we save representations in the  limit, we obtain the Askey Scheme for hypergeometric 
orthogonal polynomials. See Figure 2.}
\end{itemize}

\begin{figure}[p]
\centerline{\includegraphics[width=10cm, height=10cm]{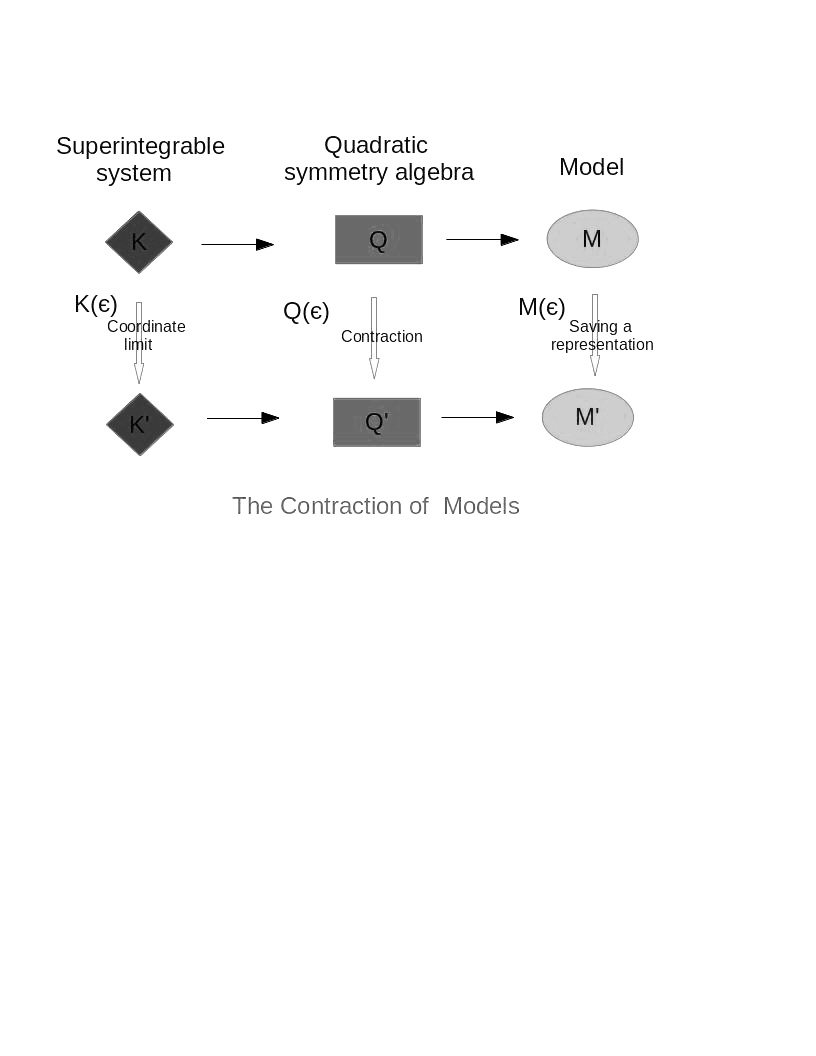}}
\caption{} 
\end{figure}

\section{Hypergeometric polynomials and the Askey scheme}  Recall,  \cite{AAR},  that the  Wilson polynomials are defined as 
\begin{equation}\label{Wilson} w_n(t^2)\equiv w_n(t^2,a,b,c,d)=(a+b)_n (a+c)_n(a+d)_n\times\end{equation}
$$ {}_4F_{3}\left(\begin{array} {llll}-n,&a+b+c+d+n-1,&a-t,&a+t \\ a+b,&a+c,& a+d\end{array};1\right)$$
$$ = (a+b)_n(a+c)_n(a+d)_n\Phi^{(a,b,c,d)}_{n}(t^2),$$ 
where $(a)_n$ is the Pochhammer symbol and ${}_4F_3(1)$ is a  hypergeometric function of unit argument. 
The polynomial $w_n(t^2)$ is symmetric in 
$a,b,c,d$.
For the finite dimensional representations the spectrum of $t^2$ is $\{(a+k)^2,\ k=0,1,\dots,m\}$ and the
 orthogonal basis eigenfunctions are Racah polynomials. In the infinite dimensional case they are Wilson polynomials. They are eigenfunctions 
for the difference operator $\tau^*\tau$ defined via
 \bea \tau&=&\frac{1}{2t}(E_t^{1/2}-E_t^{-1/2}),\nonumber\\
 \tau^*&=&\frac{1}{2t}\left[(a+t)(b+t)(c+t)(d+t)E_t^{1/2}-(a-t)(b-t)(c-t)(d-t)E_t^{-1/2}\right],\nonumber\eea 
with $E_t^AF(t)=F(t+A).$

The Askey Scheme, \cite{Koorn,KLS},  organizes the theory of hypergeometric orthogonal polynomials of one variable by exhibiting the relations such 
that each of these polynomials can be obtained as a  sequence of pointwise limits from either the Racah polynomials in the finite dimensional case or  the Wilson polynomials in the 
infinite dimensional case.
\[\lim_{\tau\to\infty}\Phi_n(\tau)=\Phi'_n.\]

The irreducible representations of $S9$ have a realization  in terms of difference operators in 1 variable \cite{KMPost1}, exactly the 
structure algebra for  the Wilson and Racah polynomials!  By 
contracting these representations to obtain the representations of the quadratic symmetry algebras of the other  superintegrable systems we 
obtain the full Askey scheme of orthogonal hypergeometric polynomials. This relationship
 ties the structure equations directly to physical phenomena. The full details of the contractions are given in \cite{KMPost13}; our contribution here is to show how these contractions were induced
in a natural and unique way from Lie algebra contractions which have clear physical and geometrical significance.  In the following we just give some examples.

\begin{figure}[p]
\centerline{\includegraphics[width=7cm, height=9cm]{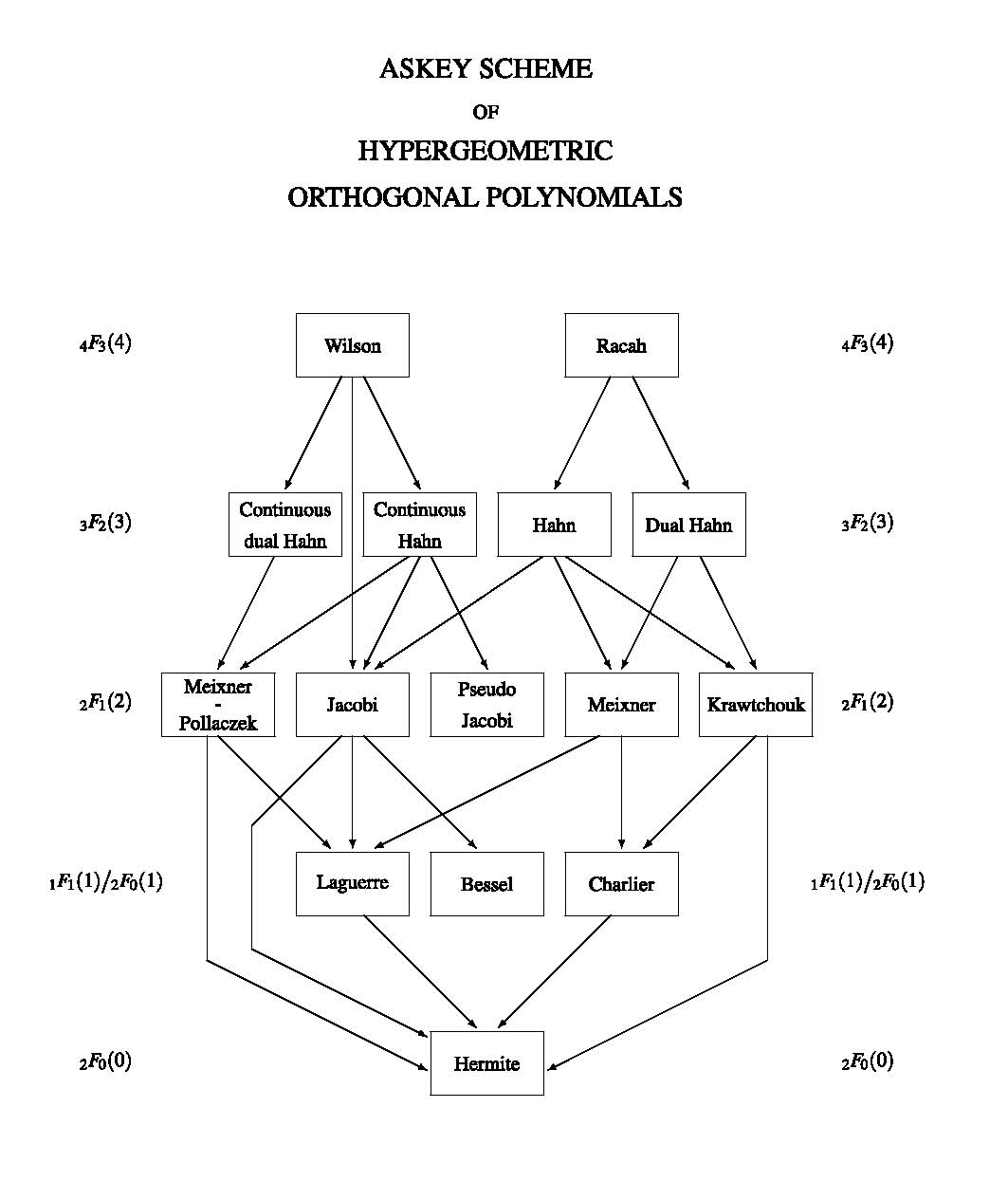}}
\caption{} 
\end{figure}

\section{The $S9$ difference operator model} 
There is no model of the irreducible representations of the quadratic algebra $S9$ in terms of differential operators but there is a difference operator model \cite{KMPost1}:
  \[   L_2f_{n,m}=(-4t^2-\frac12+ B_1^2+B_3^2)f_{n,m},\]
\[   L_3f_{n,m}=(-4\tau^*\tau -2[B_1+1][B_2+1]+\frac12)f_{n,m},\] 
\[  H=L_1+L_2+L_3+\frac34-(B_1^2+B_2^2+B_3^2)=-4(m+1)(B_1+B_2+B_3+m+1)-2(B_1B_2+B_1B_3+B_2B_3)\]
\[+\frac34-(B_1^2+B_2^2+B_3^2).\]
Here $n=0,1,\cdots,m$ if $m$ is a nonnegative integer and $n=0,1,\cdots$ otherwise.   Also 
\[   a_j=\frac14-B_j^2,\ \alpha= -(B_1+B_3+1)/{2}-m,\ \beta=(B_1+B_3+1)/{2},\]
\[ \gamma=(B_1-B_3+1)/{2},\
\delta=(B_1+B_3-1)/{2}+B_2+m+2,\]
\[  
E^AF(t)=F(t+A),\
 \tau=\frac{1}{2t}(E^{1/2}-E^{-1/2}),\]
\[
\tau^*=\frac{1}{2t}\left[(\alpha+t)(\beta+t)(\gamma+t)(\delta+t)E^{1/2}-(\alpha-t)(\beta-t)(\gamma-t)(\delta-t)E^{-1/2}\right],
\] 
\[ \ w_n(t^2)=(\alpha+\beta)_n(\alpha+\gamma)_n(\alpha+\delta)_n
{}_4F_{3}\left(\begin{array} {llll}-n,&\alpha+\beta+\gamma+\delta+n-1,&\alpha-t,&\alpha+t \\ \alpha+\beta,&\alpha+\gamma,&\alpha+\delta\end{array};1\right)\]
\[  = (\alpha+\beta)_n(\alpha+\gamma)_n(\alpha+\delta)_n\Phi^{(\alpha,\beta,\gamma,\delta)}_{n}(t^2),\quad \Phi_n\equiv f_{n,m},\]
\[   \tau^*\tau \Phi_n=n(n+\alpha+\beta+\gamma+\delta-1)\Phi_n,\]
where $(a)_n$ is the Pochhammer symbol and ${}_4F_3(1)$ is a  hypergeometric function of unit argument. The polynomial $w_n(t^2)$ is symmetric in 
$\alpha,\beta,\gamma,\delta$.
For the finite dimensional representations the spectrum of $t^2$ is $\{(\alpha+k)^2,\ k=0,1,\cdots,m\}$ and the
 orthogonal basis eigenfunctions are Racah polynomials. In the infinite dimensional case they are Wilson polynomials.

The action of $L_2$ and $L_3$ on an $L_3$ eigenbasis is
  \[ L_2f_{n,m}=-4K(n+1,n)f_{n+1,m}-4K(n,n)f_{n,m}-4K(n-1,n)f_{n-1,m}+( B_1^2+B_3^2-\frac12)f_{n,m},\]
\[ L_3f_{n,m}= -(4n^2+4n[B_1+B_2+1]+2[B_1+1][B_2+1]-\frac12)f_{n,m},\]
\[K(n+1,n)=\frac{(B_1+B_2+n+1)(n-m)(-B_3-m+n)(B_2
+n+1)}{ (B_1+B_2+2n+1)(B_1+B_2+2n+2)}
,\]
\[K(n-1,n)=\frac{n (B_1+n)(B_1+B_2+B_3+m+n+1)(B_1+B_2+m+n+1) }{(B_1+B_2+2n)(B_1+B_2+2n+1)},\]
\[ K(n,n)=[\frac{B_1+B_2+2m+1}{2}]^2-K(n+1,n) -K(n-1,n),\] 
 
We give an example showing how a contraction of one superintegrable system to another induces a similar contraction of models and recovers part of the Askey scheme.
Our example is the contraction of $S9$ to $E1$. The full scheme of limits of orthogonal polynomials is recovered through sequences of contractions of superintegrable systems,
starting from $S9$.

\medskip
\noindent {\bf Quantum system limit:} 
 \[ H_{S9}=J_1^2+J_2^2+J_3^2+\frac{a_1}{s_1^2}+\frac{a_2}{s_2^2}+\frac{a_3}{s_3^2}\]
where      $J_3=s_1\partial_{s_2}-s_2\partial_{s_1}$  and $J_2,J_3$
are obtained by cyclic permutations of the indices $1,2,3$.  
 \[ H_{E1}=\partial_x^2+\partial_y^2-\omega^2(x^2+y^2)+\frac{b_1}{x^2}+\frac{b_2}{y^2}\] 

In $S9$ we contract about the north pole of the unit sphere. Set 
\[s_1=\sqrt{\epsilon} x,\  s_2=\sqrt{\epsilon} y,\ s_3=\sqrt{1-s_1^2-s_2^2}\approx 1-\frac{\epsilon}{2}(x^2+y^2),\]
\[a'_1=b_2=a_1,\ a'_2=b_1=a_2,\ a'_3=-\omega^2=\epsilon^2 a_3,\]
 in $ H_{S9}$ to get $\epsilon({ H_{S9}}-a_3)\to  { H_{E1}}$ as $\epsilon\to 0$.

\medskip
\noindent {\bf Quadratic algebra contraction:}
\[{ L}'_1=\epsilon { L}_1,\ { L}'_2=\epsilon { L}_2,\ { L}'_3={ L}_3,\ { H}'=\epsilon({ H}-a_3)\]
\[  { R}'=\epsilon { R},\ a'_1=b_2=a_1,\ a'_2=b_1=a_2,\ a'_3=-\omega^2=\epsilon^2 a_3.\] 

\medskip
 \noindent {\bf Saving a representation:}  We set
\[ t=-x+{B_3}/{2}+(B_1+1)/{2}+m,\  B_3=\frac{\omega}{\epsilon}\to\infty \ \Longrightarrow\]  
\[f'_{n,m}= {}_3 F_2\left(\ba{lll} -n,&B_1+B_2+n+1,&-x\\ -m,& B_2+1& \ea;1\right)=Q_n(x;B_2,B_1,m)\]
where the $Q_n$ are  Hahn polynomials.  We have the  model \[L'_2f'_{n,m}= 2\omega(2x-2m-B_1-1)f'_{n,m}
=-4K'(n+1,n)f'_{n+1,m}-4K'(n,n)f'_{n,m}-4K'(n-1,n)f'_{n-1,n},\]
\[ L'_3f'_{n,m}=-\left(4n^2+4n[B_1+B_2+1]+2[B_1+1][B_2+1]-\frac12\right)f'_{n,m}=\qquad\qquad\]
\[ \left[-4(x-m)(x+B_2+1 )E^1_x+4x(x-m-B_1-1)E_x^{-1}+8x^2+4x(B_1+B_2-2m)\right.\]
\[\left.-4m(B_2+1)-2(B_2+1)(B_1+1)+\frac 12\right]f'_{n,m},\]
$$H'=L'_1+L'_2=-2\omega(2m+2+B_1+B_2).$$
Here the $K'$ are the appropriate limits of the $K$ as $B_3\to\infty$. 

\begin{figure}[p]
\centerline{\includegraphics[width=7cm]{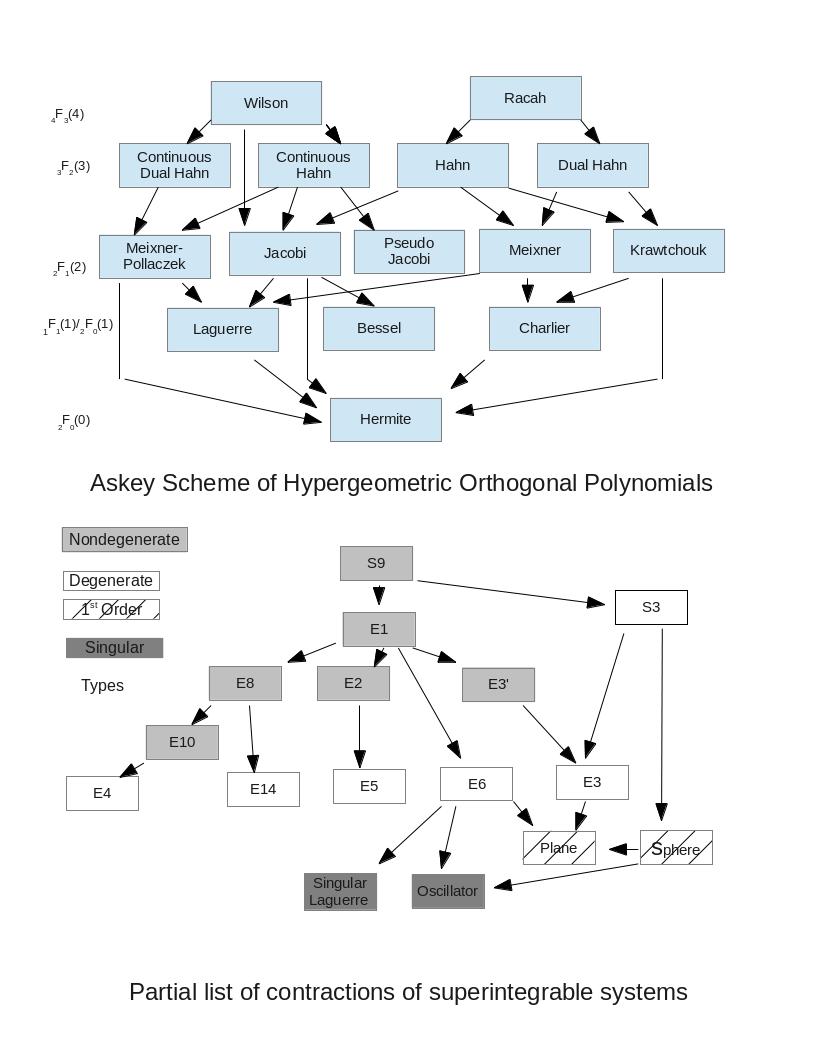}}
\caption{The Askey scheme and contractions of superintegrable systems}
\end{figure}

\begin{figure}[p]
\centerline{\includegraphics[width=7cm]{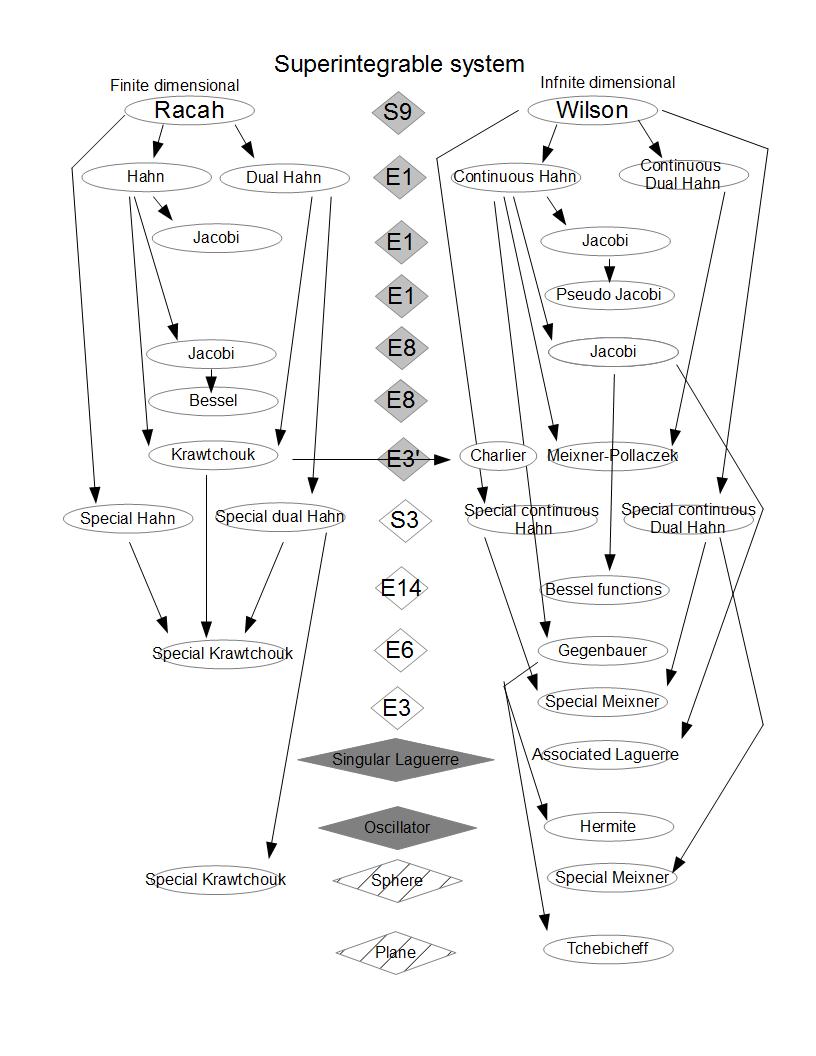}}
\caption{The Askey contraction scheme} 
\end{figure}

See Figures 3 and 4 for the contraction description of the Askey Scheme.

\section{Observations and conclusions}

 \begin{itemize}
\item Free quadratic algebras uniquely determine associated superintegrable systems with potential.
\item A contraction of a free quadratic algebra to another uniquely determines a contraction of the associated superintegrable systems.
\item For a 2D superintegrable systems on a constant curvature space these contractions can be induced by Lie algebra contractions of the underlying
Lie symmetry algebra.
\item Every 2D superintegrable system is obtained either as  a sequence of contractions from $S9$ or is St\"ackel equivalent to   a system that is so obtained.
  \item Taking contractions  step-by-step from the $S9$ model we can recover the Askey Scheme. 
However, the contraction method is more general. It applies to all special functions 
that arise from the quantum systems via separation of variables, not just polynomials of hypergeometric type, and it extends to higher dimensions \cite{KMPost11}.
The special functions arising from the models can  be described as the 
coefficients in the expansion of one separable eigenbasis for the 
original quantum system in terms of another separable eigenbasis.
The functions in the Askey Scheme are just those  hypergeometric polynomials that arise as the expansion coefficients relating two 
separable eigenbases  that are {\it both} of hypergeometric type. Thus, there are some 
contractions which do not fit in the Askey scheme since the physical system fails to have such a pair of separable eigenbases. 

\item The details of the Askey Scheme derivation can be found in \cite{KMPost13}. The origin of the complicated multiparameter contractions was   not clear in that paper. In this paper we have demonstrated that all of these  contractions were uniquely induced by the contractions of the Lie algebras $e(2,{\bf C})$, $o(3,{\bf C})$. Details will follow in \cite{KMSH}. There are only a small number of these Lie algebra contractions and their action on physical space is well known.
\item Even though 2nd order 2D nondegenerate superintegrable systems admit no group symmetry, their structure is determined completely by the underlying symmetry of constant curvature spaces.
\item To extend the method to Askey-Wilson polynomials  we would need to find  appropriate  $q$-quantum mechanical systems with $q$-symmetry algebras
and we have not yet been able to do so. 
 \end{itemize}

\ack 
This work was partially supported by a grant from the Simons Foundation (\# 208754 to Willard Miller, Jr.).
\bibliographystyle{plain}

\end{document}